# Two-stage Deep Stacked Autoencoder with Shallow Learning for Network Intrusion Detection System

[a]Nasreen Fathima, [b]Akshara Pramod, [b]Yash Srivastava, [b]Anusha Maria Thomas, [c,*]Syed Ibrahim S P, Chandran K R

[a] Research Scholar, School of Computer Science and Engineering, Vellore Institute of Technology, Chennai Campus, Tamil Nadu, India

[b] School of Electronics Engineering, Vellore Institute of Technology, Chennai Campus, 600127, Tamil Nadu, India

[c] Professor, School of Computer Science and Engineering, Vellore Institute of Technology, Chennai Campus, Tamil Nadu, India.
Email: syedibrahim.sp@vit.ac.in.

[d]Professor, Department of Information Technology, PSG College of Technology, Coimbatore, Tamilnadu, India,  Email: krc.it@psgtech.ac.in.

*Abstract*

*Sparse events, such as malign attacks in real-time network traffic, have caused big organisations an immense hike in revenue loss. This is due to the excessive growth of the network and its exposure to a plethora of people. The standard methods used to detect intrusions are not promising and have significant failure to identify new malware. Moreover, the challenges in handling high volume data with sparsity, high false positives, fewer detection rates in minor class, training time and feature engineering of the dimensionality of data has promoted deep learning to take over the task with less time and great results. The existing system needs improvement in solving real-time network traffic issues along with feature engineering. Our proposed work overcomes these challenges by giving promising results using deep-stacked autoencoders in two stages. The two-stage deep learning combines with shallow learning using the random forest for classification in the second stage. This made the model get well with the latest Canadian Institute for Cybersecurity - Intrusion Detection System 2017 (CICIDS-2017) dataset. Zero false positives with admirable detection accuracy were achieved.*

1.     **Introduction**

Information communication networks have been contributing in enhancing the performance in the means of communal and financial benefits. Moreover, malware events and breaches have become the most serious problem in the industry now. This is due to the increase in social and economic demand for information communication networks to grow the standard of people's day-to-day lives. In today's world, the maintenance and security to corroborate safety and assurance ought to be a mandatory aspect of every organization. To do so, the data communication and any other network have to be shielded from such threats or attacks. Network security is becoming increasingly important as the usage of the Internet of things continues to grow in a plethora of different specializations and software. Many organizations use conventional security tools to defend against web attacks. Anti-spam techniques, firewalls, anti-viruses, etc., are some of them. These tools fail to identify unknown and advanced persistent threats. Hence, an Intrusion detection system plays a vital role in securing network and data entities in a system or a computer.

An Intrusion detection system acts as a tool to monitor malign behaviours in a system. It makes sure to help the system or network in defending against the latest exploits. In general an intrusion detection system can be classified into two types, they are network-based intrusion detection system and host-based intrusion detection system [1]. The network-based intrusion detection system is being used to monitor the network and identify any malign events [10]. It serves as a defense tool to identify and send alerts against these threats and attacks. A host-based intrusion detection [24] system detects suspicious events or software in a host system alone. It doesn't monitor the network traffic in it. The techniques used to detect these systems are signature-based detection and anomaly-based detection [10]. Signature-based detection reveals a known pattern of malicious events, whereas anomaly-based detection identifies new malware. Since signature-based detection has no potential to detect unknown attacks, it cannot be a promising technique to apply to a real-world system. Anomaly-based detection overcomes this challenge and is assured to be able to identify unknown damages. However, many anomaly detection methods have a high false alarm rate.

In recent years, researchers show high interest in introducing machine learning techniques.Some of the widely accepted algorithms are  Support Vector Machine, KNearest Neighbour, Artificial Neural Network, Random Forest, Decision Tree and Linear Regression [4].  These machine learning techniques work well and are able to predict the malwares better than the traditional methods. However, they are  not efficient to predict unknown attacks in the real world environment and hence deep learning helps by dominating in solving these problems.

Deep learning is a part of machine learning. It has created a great impact in terms of decision making and predictive analytics. It has become a research hotspot in solving a plethora of real time problems. The need for higher detection rates and less false positives are in demand, rather than accuracy. Hence, to handle the challenges in handling large scale traffic, deep learning techniques ought to be giving astonishing results.

Recently, researchers are focusing on autoencoders [17],[18],[28] in anomaly based detection. An autoencoder is a kind of artificial intelligence which relies on artificial neural networks. It is used to learn valuable information by encoding data in an unsupervised way. The target of an autoencoder is to set up an encoding architecture with the available data to reduce the dimension. It achieves the target by training the neural network in terms of removing noises. It also reconstructs the data after learning and tries to generate a low dimensional data representation of it.

Shone et. al [28] proposed a non-symmetric deep stacked autoencoder for NIDS. It is different from a basic autoencoder which produces dimensionality reduction. This work combines RF for classification by merging both deep learning and shallow learning to perform analytics. This shows better performance with less training time. Two stage deep learning autoencoder [17] was proposed based on a stacked autoencoder to train in both the stages with softmax classifier at the output layer. They have taken probability in the first stage to feed the output value as an additional feature for the next stage. Even Though this method shows a better prediction accuracy, this method used only 5 features in the first stage. This proposed work contributes to handling feature extraction and improving the accuracy in classification by applying deep stacked autoencoders and shallow learning with all the features in every stage of deep learning. So, the proposed system provides more generalizability and explainability to the model.

The recent CICIDS-2017 dataset is applied to assess the model. The developed architecture overcomes the inefficiency of preceding architectures on behalf of feature importance with two stage deep learning. Hence the execution of the proposed model is impressive.

This paper is structured as follows. Section II looks over related works on network-based IDS, which narrows down the focus towards deep learning architectures for feature extraction. In Section III, we discuss an overview of autoencoder architectures used in the experiments and present the two stage deep learning model which extracts the features to do classification using RF classifier acting as a shallow learner. In Section IV, we demonstrate the improvement by using the features processed from deep stacked autoencoders and the effectiveness of the two stage DSAE-RF. Finally, we provide a conclusion in Section V.

2. **Related Work**

During the last decade, many methods, models, algorithms and approaches have been introduced to find the solution for network intrusion detection problems using machine learning and deep learning techniques. Different types of methods are reported in later literature to classify malware attacks in a network [4]. Janarthan and Zargiri [17] applied a Random Forest classifier to classify intrusion attacks on the UNSW-NB15 dataset by extracting only five features. This model had a comparatively lesser accuracy, nearly about 81.61% which was believed to be improved further. Even so, the traditional methods seem to be inefficient when deployed for a real network traffic analysis. The main challenges in this field are inability to handle huge data size, less detection accuracy of attacks, inefficient for dynamic analysis, and inability to adapt to the nature of the products for which they are used. [28]

In addition to the machine learning algorithms, deep learning strategies are a part of it, which has shown promising results in recent times in the field of network intrusion detection. According to most of the observations and studies, deep learning has persuaded us to be more effective than classical techniques. The deep learning architectures can also be designed to perform supervised and unsupervised feature learning. Some of the models that have been implemented in the preceding are Restricted Boltzmann Machines (RBMs), Autoencoder (AE), Deep Belief Networks (DBNs), Deep Neural Networks (DNNs), Convolutional Neural Networks (CNN), Long Short Term Memory Recurrent Neural Network (LSTM RNN) and some other variations on these foundational methods.

Alrawashdeh and Purdy [12] presented a deep learning method relying on a framework DBN of RBMs. This method involves four hidden layers for feature diminution. Syed et al. [3], have experimented with feature selection as a step using the traditional KDDcup99 dataset to reduce the possibility of overfitting and data redundancy. The results were comparatively better than the preceding work, but still these models were not effective to deal with large datasets and overcome real time challenges such as false alarm rate. The fine-tuning phase consisted of updation of weights of the DBN, and on the other hand a Logistic Regression Classifier was used for behavioural classification. This procedure was also applied on the KDD99 dataset. The accuracy turned out to be 97.9%. In their work, Erfani et al. [13] have discovered an unique combination of DBNs with a sequential single-class SVM in the field of NIDS and has been tested on many deep-rooted datasets.

Yin et al. [18] innovated a deep learning procedure for network intrusion detection which is solely incorporated using Recurrent Neural Networks (RNNs). This process was implemented to one of the benchmark dataset, NSL-KDD and suggested that deep learning techniques are more effective and outperform traditional machine learning approaches in case of NIDS. Lately, Nguyen et al. [19] presented a framework that targeted detection of network attacks in the mobile in a cloud. This method was brought out from the core concept of principal component analysis (PCA) and Gaussian-binary restricted Boltzmann machine (GRBM). Although, this approach has a little unclear testing methodology which is insufficient to enable comparative benchmarking. Jihyun Kim proposed in one of his works that LSTM RNN can be used on a KDD99Cup dataset to implement an IDS classifier. This paper compared the traditional RNN model and the LSTM RNN model and suggested that LSTM RNN provided an improved efficiency of about 96.93% [20]. Nawir et al. [23], have implemented classification models dependent on the Average One Dependence Estimator algorithm (AODE). The model is based on multi-class classification and is tested using the UNSW-NB15 dataset. The metrics numbers were pretty better comparatively. Javaid et al. [25] proposed a deep learning-based algorithm on Deep Neural Network for outlier forecast. His studies suggested that for flow-based anomaly detection, a deep learning technique relied on DNN is more effective if a network is a software-defined network. A deep learning approach on the NSL-KDD99 dataset based on STL was introduced by Tang et al. This work suggested that deep learning outperforms previous studies according to accuracy.

Later, Wang's proposal involved a deep learning procedure based on stacked autoencoders and remarkably attained a very high accuracy and performance. K. Wu et al. proposed a CNN model for classifying network intrusion detections. This proposed work converted the raw traffic vector format into an image data format to tackle real-life data. It was implemented on the NSL-KDD99 dataset which was featured from the KDD99Cup dataset. However, this model's accuracy could not cross 81%, which showcased a path for further improvement.

Deep learning algorithms such as stacked autoencoder (SAE), restricted Boltzmann machines (RBMs) in combination with supervised convolutional neural networks (CNNs) are used in preceding works. Even though the numerical values of parameters get reduced while implementing CNNs through strategies of shared weights and sparse connectivity, these methods need a considerable amount of categorical data to be fed as an input which is relatively expensive and so they used supervised learning [30]. The challenges faced in the majority of research are low detection rates of each class, high false alarm rate, low precision and no zero day attack is achieved.

In preceding work, the encoding layers concentrate on the quantity of information rather than quality. Nathan et al [28] presented a novel approach on deep learning for unsupervised feature engineering. Their work represents a remarkable result that is efficient to overcome dimensionality problems rather than feature extraction. Loss of features is not appreciable to handle network traffic. Hence, Farrukh et al [17] has developed a two stage autocoder, where the outcome of the first stage turns out to be an input for the next stages with the desired feature delineation and reduction. The framework extracts the features for categorizing normal and malign network bottlenecks with a probability result in the first stage. This result obtained from the first stage is used as a supplementary feature in the next phase to extract the feature anomalous network traffic. The overfitting and feature reduction issues towards normal traffic and malign traffic are addressed in this work. This not only confirms the detection of the normal traffic but also classifies other sorts of damages. This subject study's motive is to model a two-stage feature engineering derived from deep learning to boost accuracy and reduce the false alarm rates. However, the challenges faced in this work are feature extraction, high false alarm rate and inefficiency for real world network traffic without loss of data. This is due to the lagging of potentiality to handle large scale traffic of the latest updated attack dataset such as CICIDS-2017 and feature engineering without data loss.

This study is developed in a more sophisticated way to learn more useful feature representation. It aims at involving all the features provided in the dataset so that the model can tackle any real-life data in the future. Similar to a work proposed by Farrukh Aslam Khan, our model is structured as a Two-Stage Deep Learning Model. His work only includes five abstract features. Our work has tried to bend more towards real-life scenarios and therefore suggests classification based on the latest updated CICIDS-2017 dataset, with updated attacks which resembles real time data. There is no data loss as we have used all the features extracted. This study is more promising with the best accuracy, precision, recall, and performance.

Methodology

*2.1. Deep Stacked Autoencoder*

Stacked autoencoder is a stack of autoencoders as hidden layers in the neural network. Stacked autoencoder is an unsupervised learning method that applies a back propagation technique to project the output value. This neural network improves accuracy in deep learning with noisy autoencoders embedded in the layers. Stacked autoencoders generally include three steps as follows.

Autoencoder is trained using the input and obtain the acquired features. These features act as an input for the next layer, and the same process is performed again until the training ends. Back-propagation algorithm (BP) is applied to reduce the cost function if the hidden layer is trained and renovate the weights by labelling training data to attain desired performance.

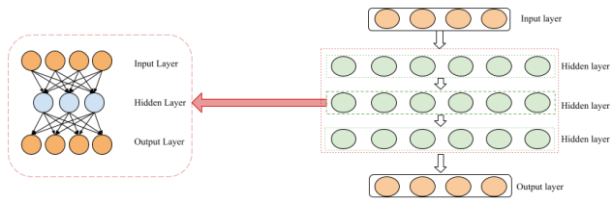

Fig. 1. Fundamental Architecture of Deep Stacked Autoencoder

Let the input layer be represented by x. Let the hidden layer have m nodes, which contain abstract compressed data d. Let the output layer's values be given by y.

$$\text{for } i \in n \quad x_i \sim y_i \qquad (1)$$

Let w, b1 represent the weight matrix and the bias vector respectively.

We also need an activation function (represented by g). In our case, we have selected the Rectified Linear Unit (ReLu) activation function.

$$\text{Now,} \quad d = g(xw + b_1) \qquad (2)$$

The decoding process of the autoencoder is given by

$$y = g(dw^T + b_2) \qquad (3)$$

where b2 and wT represent the bias vector and the weight matrix respectively.

When we obtain the values of wT and b2, then the classification of the labelled data (x,y') is performed using a supervised learning algorithm. Back-propagation algorithm then adjusts the values of wT and b2 explicitly. The new features d, vector y' are tuned by using SoftMax regression classifier.

*2.2. Softmax Classifier*

SoftMax is a neural network layer that is implemented just before the output panel. Number of nodes in this panel must be identical to the output panel. SoftMax is logistic regression generalized, used for multi-class classification. Most multi-layer neural networks finish in the second last layer whose outcomes real-valued scores don't fit well. Because it is difficult to proceed with this, SoftMax transforms the scores to a standardized probability, which can be shown to the user or proceed as an input to the different system. Hence, it is normal to add a SoftMax function as the innermost layer of the neural network.

SoftMax Formula

We use scipy.special.softmax

softmax(x) = np.exp(x)/sum(np.exp(x))

for a vector x = {x0, x1, x2, . . . . xn-1 }, the formula for SoftMax function is given by

$$\sigma(x_j) = \frac{e^x}{\left[\sum_j e^{x_k}\right]} \qquad (4)$$

*2.3. Random Forest Classifier*

Random forest classifiers can be either classification or regression analysis. It is a supervised learning technique. Mostly, the random forest algorithm is applied for classification purposes. The fact that forest is full of trees, Random forest is a combination of many decision trees like a dynamic model that acts as an ensemble model to reduce the high variance. This ensemble method has leaf nodes as the majority class or as an average for regression problems.

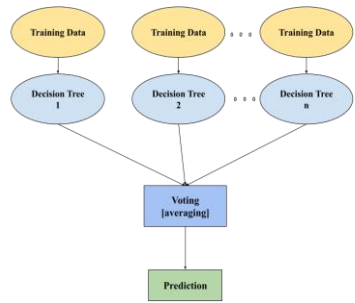

Fig.2. Random forest Architecture

3. **Proposed framework**

In our proposed model, we apply a deep learning model in two stages. Considering merits such as performance and speed in real time classification situations, we decided to involve a DNN based procedure for the same. We start by reading the latest CICIDS-2017 dataset and perform appropriate pre-processing techniques. When we begin the 1st stage of our model, we pass in all the columns as features and classify the network traffic into two categories- normal and attack. We also generate a probability score for the two classes.

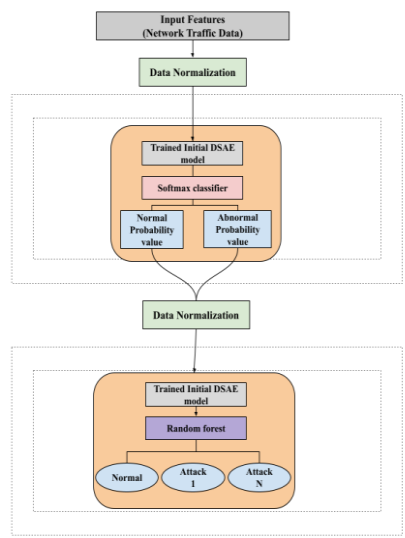

Fig.3. Proposed Two stage Deep learning model

*Algorithm*

1. begin
2. import and install all necessary packages and libraries
3. read the CICIDS-2017 dataset
4. features —> all columns, classes —> label
5. training_data, testing_data, training_classes, testing_classes —> train_test_split(features,classes)
6. training_outcomes, testing_outcomes —> unique labels in training and testing classes
7. classify labels into two classes —> normal and attack
8. scale and encode training and testing data-frames
9. apply deep stacked autoencoder
10. apply SoftMax classifier
11. add the obtained probability values to the testing data-frame
12. feed the output as an additional feature for the next stage
13. classify using random forest as a shallow learning
14. repeat steps 9 and 13
15. end

We use this probability value in addition with other features to train the 2nd stage classifying them as normal and attack. Both the stages' logic is similar and combines a Deep Stacked Autoencoder consisting of two hidden layers. The output of the autoencoder in the first stage is fed to a SoftMax layer classifier for generating the probability score. Then the output is fed to a Random forest classifier as a shallow learning in the second stage which produces tremendous improvement in the results.

**4.     Experiments and Evaluation**

*4.1. Performance Measures*

To validate the working of the proposed model, we use the very same validation metrics chosen in many researches in this field. All these metrics are evaluated with the help of the following calculations:

$$Overall\ Accuracy\ (Acc) = \left(\frac{TP+TN}{TP+TN+FP+FN}\right) \quad (5)$$

$$Precision\ (P) = \left(\frac{TP}{TP+FP}\right) \quad (6)$$

$$Recall\ (R) = \left(\frac{TP}{TP+FN}\right) \quad (7)$$

$$F_1 Score = 2\left(\frac{P \times R}{P+FP}\right) \quad (8)$$

Here, the term TP stands for true positives, TN stands for true negatives, FP stands for false positives and FN stands for false negatives. Moreover, the presented model is used for classification and is measured to reveal the model's capability for real-time execution.

*4.2. Dataset Description*

There are numerous publicly accessible dataset for the evaluation of NIDs. But, they do not replicate the real time network traffic. A demanding task in NIDS is to assess the degree of pragmatism of these datasets. So, researchers use several evaluation metrics that calculate the angle of pragmatism, and implies more appealing datasets that contain measurable rapid patterns of benign and malign traffics. This is of vital importance to develop a network intrusion detection system in attaining reliability and plausibility. The dataset used in this work is CICIDS-2017 dataset. This is a benchmark dataset used to analyze in accordance with the accuracy and detection rate to validate the model's efficiency.

CICIDS-2017 dataset is the most updated one which can be used to evaluate NIDS. The malign and benign patterns of behaviour in the other benchmark datasets are outdated, there is no influence to the current scenarios of network traffics. So the performance estimations are misleading. Hence we chose CICIDS-2017 dataset to train the model. The higher accuracy rates in existing models are indicating the reasons behind it.

CICIDS2017 dataset has benign and the most trending cyber attacks, which replicates the true world network traffic data. The class names are already labelled as normal or attack in the dataset. There are three groups of attributes in the CICIDS-2017 dataset. They are network traffic, content and primary attributes. The traffic attributes are categorized into two types one is host feature and the other one is service feature. It is estimated by testing the connections created between every two-seconds [17]. The malicious patterns are recognized in content features. And the primary attributes are abstracted from the TCP/IP links. The attacks such as Brute Force FTP, Brute Force SSH, DoS, Heartbleed, Web Attack, Infiltration, Botnet and DDoS are included.

*4.3. Data pre-processing*

In this stage, we convert the raw data into data representation suitable for our application. We observe that in the CICIDS-2017 dataset, we have unrecognized character in the class label. We changed the unrecognized character in the class label. We then combined all the dataset. We converted these values in the dataset to numeric values because the proposed two-stage deep learning model only works with numeric features. So, we take all the attributes of features and convert them to numeric integer values. This is achieved by defining a function to encode all the non-numeric attributes to numeric values.

4.4. Feature normalisation

To ensure that the range of the numeric data doesn't affect our model, we perform feature normalization. For a given range of values x1, x2, x3,…, xn stored in an array X, the normalized values u1, u2, u3, …, un are given by

$$u_i = \frac{x_i - max(X)}{(max(X) - min(X))}$$

To do so we utilize the MinMaxScaler package from sklearn. This scaler is applied to all the numeric values to give normalized values.

5. **Results and Comparisons**

Subsequent to the implementation of the proposed TSDL model using Python, we experimented on the carefully chosen CICIDS-2017 dataset with a 10-fold cross-validation procedure. In this procedure, the given dataset is split into ten chunks, where one of the chunks is employed as a testing set and the remaining are employed as a training set. This process is reiterated 10 times which gives us the standard results yielded into a single rough calculation. The privilege of this estimation is that all the studies are working for both training and testing, and each study can be made use of only once for testing and training the model. Here, we plan to show the results at every single stage of the model. The outcome metrics are revealed on both first and second decision stages.

Basically, the first decision stage is in charge of the classification, on the other hand the second decision stage is responsible for performing multi-class classification. With this, we can infer that the proposed TSDL model is an adaptable model with two alternatives that can be exercised as per the user's requests.

The given CICIDS-2017 dataset consists of 80 features which are normalized and passed into the next stage for initial training using the DSAE model. They are further classified into normal and attack categories using the SoftMax classifier. The accuracy at this stage is 85%. The classification of benign and malign traffic at the first stage is typically designed to estimate on the forecast of the outcome using SoftMax classifier, ranging from 0 to 1. This predicted probability output is processed as a supplementary feature for the next stage. The additional features appended to the 80 features arising from the first stage are all together trained using the final DSAE model. These features are again used to classify benign and malign classes by applying Random forest algorithm to classify the classes in the second decision stage. The accuracy of the model is 87%.

**Table 1.** Results comparisons between deep stacked autoencoder and two stage deep learning with shallow learning

| CICIDS-2017 dataset | F1 score | Recall | Precision | Accuracy % |
|---|---|---|---|---|
| Deep Stacked AE | 0.55 | 0.56 | 0.50 | 0.63 |
| Stacked AE with Softmax classifier | 0.72 | 0.72 | 0.71 | 0.71 |
| Stacked AE with RF classifier | 0.77 | 0.76 | 0.74 | 0.71 |
| Two stage AE with Softmax classifier | 0.79 | 0.77 | 0.72 | 0.73 |
| Two stage DSAE with Random forest | 0.81 | 0.79 | 0.82 | 0.87 |

6. **Conclusion**

This work offers an interesting novel framework using deep learning for detecting intrusions from network traffic data The proposed framework comprises two stages: two hidden layers, softmax classifier in first stage and random forest in second stage. In a semi supervised manner, the deep learning model is trained. The next step is to pre-train each hidden layer independently on a huge collection of unlabelled data in an unsupervised way. After that the features are adjusted by applying labeled network features.

The first step is to train the model and classify whether it is a normal state or attack state. The user can choose the deep learning model at this phase only. The final decision layer is employed to detect normal and various kinds of attacks. The proposed model helps the network intrusion system to work accurately and categorize different kinds of attacks. This fundamental experiment was executed on the CICIDS-2017 dataset. In the proposed experiment, we followed two steps: feature normalization and data preprocessing to make them more amenable in detection, and the other was multi-class classification.

This model excelled in all evaluation measures like accuracy, precision, recall and F1 score. We plan to hybridize this deep learning model to handle class imbalance problems in the near future with advanced GAN to make most effective use of our proposed network intrusion detection system.